\documentstyle[aps,prl,epsfig]{revtex}
\draft

\begin{document}
\twocolumn[
\hsize\textwidth\columnwidth\hsize\csname@twocolumnfalse\endcsname

\title{Andreev scattering in the asymmetric ladder 
with preformed bosonic pairs}
\author{Karyn Le Hur}
\address
{D\'epartement de Physique Th\'eorique, Universit\'e de Gen\`eve,
CH-1211, Gen\`eve 4, Switzerland}
\date{\today}
\maketitle

\begin{abstract}
We discuss the phase coherence which emanates from
the ladder-like proximity effect between a ``weak superconductor'' with 
preformed bosonic pairs (here, a single-chain Luther-Emery liquid with 
superconducting correlations that decay approximately 
as $x^{-1}$) and a Fermi gas with 
unpaired fermions. Carefully studying tunneling mechanism(s), we show
that the boson-mediated Cooper pairing
between remaining unpaired electrons results in a quasi long-range 
superconductivity: Superconducting correlations decay very 
slowly as $x^{-\eta}$ with $\eta\approx 1/2$. 
This process is reminiscent of the coupling of fermions
to preformed bosonic pairs introduced
in the context of high-$T_c$ cuprates.
\end{abstract}

\vfill
 \twocolumn
\vskip.5pc ]
\narrowtext

Surprisingly, the excitation spectrum in underdoped 
high-$T_c$ cuprates 
exhibits a pseudogap at a temperature $T^*$ far above the 
superconducting temperature $T_c$. 
Experimental results suggest that
the gap formation is due to the pairing of electrons in the corners of
the Fermi surface into preformed bosonic pairs (with a small
spectral weight), whereas electrons
in the ``diagonals'' would still remain unpaired (See Fig. 1)\cite{Exp-Rev}. 
Theoretically, the truncation of the Fermi surface
has been predicted by Rice {\it et al.} from 
Renormalization Group arguments in two dimensions\cite{Maurice-2D} and
from the behavior of lightly doped ladders\cite{Maurice-ladder}.
This has been also emphasized by Lee and Wen based on gauge theory calculation
\cite{WL}.
The superconductivity which emanates from the Bose condensation of these
preformed pairs coexisting with unpaired fermions has been examined 
phenomenologically by
Geshkenbein, Ioffe and Larkin\cite{GIL}.
Below, we introduce a simple {\it asymmetric} two-leg ladder
system which allows to rigorously exemplify
the (almost) Bose condensation of preformed bosonic pairs in the 
vicinity of unpaired fermions.

We discuss the quasi long-range superconductivity emerging 
at $T_c$ from the ladder-like proximity effect between a one-dimensional (1D)
weak superconductor --- which already displays a spin gap and preformed
Cooper pairs below $T^*$ --- and a 1D Fermi gas. 
\begin{figure}[ht]
\centerline{\epsfig{file=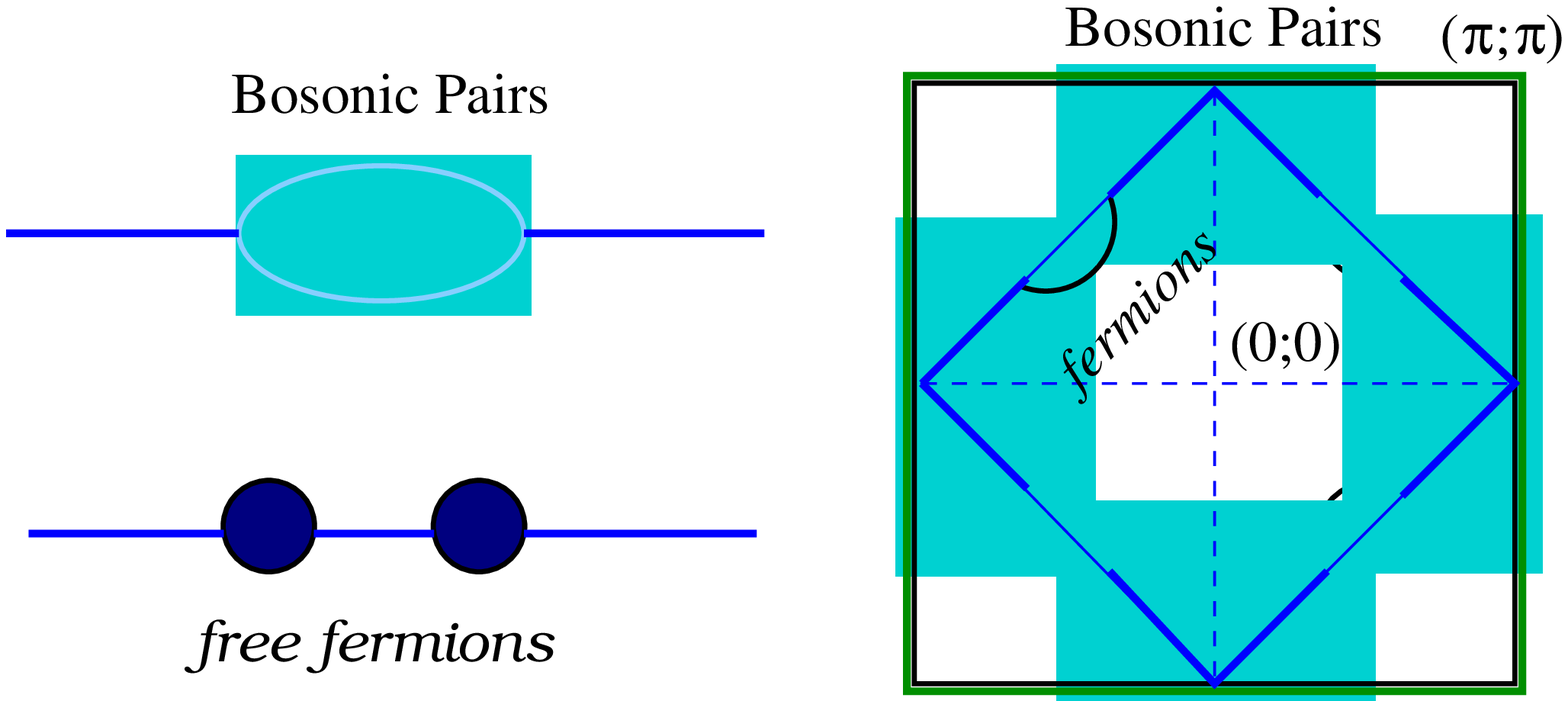,angle=0.0,height=4cm,width=9.1cm}}
\vskip 0.5cm
\caption{A ladder system which allows to investigate the Bose condensation
of preformed pairs coexisting with unpaired fermions (discussed
for high-$T_c$ cuprates): {\it A single-chain 
Luther-Emery liquid weakly-coupled to a Fermi gas}.}
\end{figure}

To keep the discussion as clear as possible, here we
assume that the weak superconductivity emerges from the single-chain 
Hubbard model with small attractive interactions, producing
the usual Luther-Emery liquid\cite{LE}. Above $T_c$ 
(but below $T^*$), the 
superconducting correlations are weak in the sense that they still decay 
approximately as the charge
density wave correlations, i.e., roughly as $x^{-1}$.
Such a prototype system has some similarities to the one
introduced by Emery, Kivelson and Zachar
in another context [a metallic stripe coupled to an active
spin-gapped doped Mott insulator]\cite{EK}. 
Following the methodology developed in Ref.\cite{KLH}, we build 
the relevant tunneling process(es) between the two chains. Here, this 
corresponds to an Andreev scattering mechanism\cite{deGennes}.
Then, we show how 
the boson-induced Cooper pairing between unpaired fermions
results below $T_c\propto 1/\{T^*\}^2$ 
both in a spinon-pairing (spin gap) in the Fermi gas\cite{EK}
and in a quasi long-range phase coherence. 
In contrast with Ref.\cite{EK} (page 6131), we report that the superconducting 
correlations now decay as $x^{-\eta}$ with 
$\eta\approx 1/2$, whereas the charge density wave 
correlations at the wave vector $q=2k_F$
arise only at short distances. The system 
behaves as a conventional two-leg ladder\cite{Fisher,Tsvelik,KLH}.
\vskip 0.1cm
For temperatures $T\gg T^*$, the kinetic energy for fermions 
takes the standard form 
${\cal H}_{kin}={\cal H}_o+{\cal H}_{\perp}$ where
\begin{eqnarray}
{\cal H}_o&=&-t\sum_{j,\alpha} 
\psi_{j\alpha}^{\dagger}(x+1)\psi_{j\alpha}(x)
    +{\mathrm H.c.},
  \nonumber\\
{\cal H}_{\perp}&=&-t_{\perp}\sum_{\alpha}
\psi_{2\alpha}^{\dagger}(x)
\psi_{1\alpha}(x)+{\mathrm H.c.}
\end{eqnarray}
Here $j=(1,2)$ denote the chains and
$\alpha=(\uparrow,\downarrow)$ the spin projections of an
electron. The bare short-distance cutoff is equal to
$a=1$. The longitudinal and transverse hopping amplitudes are respectively
$t$ and $t_{\perp}$.  

 {\it We consider the case of a weak-proximity effect between the two 
chains, i.e., we 
start with} ${\mathit t_{\perp}/T^*\ll 1}$. For weak interactions in
chain 1, our
temperature $T^*$ will be already very small compared to the Fermi energy 
$E_F\sim t$.
\vskip 0.1cm
Focusing on electronic states near the Fermi points, we expand 
$\psi_{j\alpha}(x)=\psi_{+j\alpha}(x)e^{ik_Fx}+
\psi_{-j\alpha}(x)e^{-ik_Fx}$; Below $p=\pm$ denote respectively right and 
left excitations. 
Unlike in Ref.\cite{EK}, the two chains are here equally doped.

{\bf Weak superconductor at} ${\mathbf T^*.}$
Let us first consider that $t_{\perp}=0$.
As emphasized in the introduction, we start with an asymmetric situation 
where in chain 2 electrons are completely ``free''  whereas electrons of
chain 1 are subjected to a weak attractive Hubbard interaction of the
form: ${\cal H}_{1}=
g_c J_{+11}J_{-11}-g_s {\mathbf J}_{+11}{\mathbf J}_{-11}$.
We have rewritten the Hubbard interaction in chain 1 as a function
of the usual charge and spin currents [For the
definitions: See Ref.\cite{KLH}, Eq. (5)].
Note that
$g_c$ and $g_s$ describe charge- and spin backscatterings respectively.

The bare interactions are of the order of the {\it negative} Hubbard coupling,
$U<0$. 

Bosonizing the 1D Fermi fields, we can write\cite{Tsvelik}
\begin{equation}
\label{bos}
\psi_{pj\alpha}\approx\exp\hbox{\Large(}i\sqrt{\frac{\pi}{2}}\left[p(\Phi_{jc}+\alpha\Phi_{js})
    -(\theta_{jc}+\alpha\theta_{js})\right]\hbox{\Large )},
\end{equation}
$\alpha=\pm$ for spin up and spin down respectively. 
The electron spectrum
exhibits spin-charge separation. The charge part of the Hamiltonian
results in the Luttinger Gaussian model\cite{velocity}:
\begin{equation}
H_{oj}^c=\frac{v_F}{2\pi}\int dx\ \frac{1}{K_{jc}}{(\rho_{jc}-\rho_o)}^2
+K_{jc}{(\nabla\theta_{jc})}^2.
\end{equation}
Here $v_F=2t\sin k_F$ denotes the Fermi velocity,
$\partial_x\Phi_{jc}=(\rho_{jc}-\rho_o)$ measures
fluctuations of charge density in each chain, and $\nabla\theta_{jc}$ is the
conjugate momentum to $\Phi_{jc}$. 
The free spin Hamiltonians have the same forms as the charge
Hamiltonian of Eq.(3) replacing $\Phi_{jc}$ by 
$\Phi_{js}$. The Luttinger exponents are,
\begin{eqnarray}
K_{1c}&=&1-g_c/\pi v_F>1\qquad K_{2c}=1\\ \nonumber
K_{1s}&=&1+g_s/\pi v_F<1\qquad K_{2s}=1.
\end{eqnarray}
\vskip 0.1cm
For attractive interactions in chain 1 $[K_{1s}<1]$, 
the spin backscattering term $g_s$
provides another (relevant) 
contribution to the low-energy physics. 
Indeed, this produces the Sine-Gordon model\cite{LE,Tsvelik}
\begin{equation}
H_{1}^s=H_{o1}^s-g_s\int dx\ \cos[\sqrt{8\pi}\Phi_{1s}](x).
\end{equation}
The coupling $g_s$ is known to be strongly renormalized at
the temperature scale $T^*=E_F\exp(-\pi v_F/g_s)$. The spin field 
gets pinned at the classical value $\Phi_{1s}\approx 0$. The
resulting spin gap is of the order of $T^*$ (See Ref.\cite{Tsvelik}
page 76). This produces the growth of the
superconducting (SC) and charge density wave (CDW) 
correlations in chain 1. These are given by the
operators
\begin{eqnarray}
\label{cor}
{\cal O}_{SC}^1&=&\psi_{+1\uparrow}\psi_{-1\downarrow}+\psi_{-1\uparrow}
\psi_{+1\downarrow}\\ \nonumber
&\propto& \exp(-i\sqrt{2\pi}\theta_{1c})\cos(\sqrt{2\pi}\Phi_{1s})
\end{eqnarray}
and
\begin{eqnarray}
{\cal O}_{CDW}^1&=&\psi^{\dagger}_{+1\uparrow}\psi_{-1\uparrow}+\psi
^{\dagger}_{+1\downarrow}
\psi_{-1\downarrow}\\ \nonumber
&\propto& \exp(-i\sqrt{2\pi}\Phi_{1c})\cos(\sqrt{2\pi}\Phi_{1s}).
\end{eqnarray}
The operator ${\cal O}^{1\dagger}_{SC}$ 
describes the {\it preformed bosonic pairs} with charge $Q=2e$ and spin
$S=0$. Using the fact that there is no fluctuation of the
spin field $\Phi_{1s}\approx 0$ for temperatures 
smaller than $T^*$, one finds for the charge-density correlation function
\begin{equation}
<{\cal O}^{1\dagger}_{CDW}(x){\cal O}^1_{CDW}(0)>\ \propto x^{-K_{1c}}
\end{equation}
and for the SC pairing correlation function
\begin{equation}
<{\cal O}^{1\dagger}_{SC}(x){\cal O}^1_{SC}(0)>\ \propto x^{-1/K_{1c}}.
\end{equation}
As expected, since $K_{1c}>1$ the superconducting correlations in
chain 1 are (slightly) more important than the CDW correlations
for $T<T^*$ (See Fig. 2). Here, pairing
of spinons is somewhat equivalent to pairing of electrons inducing prevalent
bosonic pairs in chain 1.

As long as the
chain 2 remains weakly ``coupled'' to chain 1 (i.e., for $T>T_c$, see
discussion below), all its density-density
correlation functions decay as $1/x^2$ as for any noninteracting 1D electron
gas.

{\bf Fate of transverse electron motion.}
Now, we examine the influence of the finite (bare)
transverse hopping amplitude. 
Again, we assume that the spin backscattering term $g_s$
is flowing {\it first} to strong couplings 
(at the temperature $T^*$;
$g_s(T^*)\approx t$) which 
corresponds to very small bare values of $t_{\perp}$
$(t_{\perp}\ll T^*)$. The bosonic form of the term $t_{\perp}$ (at $q=0$) 
is given, e.g., in Ref.\cite{KLH}.

We find that $t_{\perp}$ evolves according to
\begin{equation}
\label{flow}
\frac{d\ln t_{\perp}}{dl}
=\frac{7}{4}-\frac{1}{4}(K_c^-+\frac{1}{K_c^-})
-\frac{1}{8}(K_{1s}+\frac{1}{K_{1s}})\cdot
\end{equation}
The $l$ describes the renormalization of the short-distance cutoff 
$a(l)=\exp l$.
The renormalization procedure is stopped at lengths $a(l)$
comparable to the thermal length $v_F/T$, which means $l=\ln(E_F/T)$. We have
combined the charge boson fields in the two chains into a symmetric ``+''
and antisymmetric ``-'' part: $\Phi_c^{\pm}=[\Phi_{1c}\pm\Phi_{2c}]/\sqrt{2}$
and similarly for the conjugate momenta. 
The resulting Luttinger exponents obey
\begin{equation}
K_{c}^{\pm}=\sqrt{K_{1c}K_{2c}}\approx 1-\frac{g_c}{2\pi v_F}>1.
\end{equation}
Furthermore, we obtain
\begin{equation}
\label{13}
\frac{dK_{1s}}{d\ln(\frac{E_F}{T})} = -\frac{1}{2}(g_sK_{1s})^2.
\end{equation}

As long as $T>T^*$, integrating Eq.(\ref{flow}) produces
a linear growth of $t_{\perp}(T)$ [i.e., $g_s(T)\ll t$ and $K_{1s}\approx 1$]
\begin{equation}
\label{tt}
t_{\perp}(T)\approx t_{\perp}E_F/T\ll t.
\end{equation}
Second, from Eq.(\ref{13}) 
we deduce that the explicit divergence of the spin backstattering $g_s$
at the opening of the spin gap (i.e., for $T=T^*$) 
results formally in $K_{1s}\{T<T^*\}=0$. Using Eq.(\ref{flow}), this also 
implies
\begin{equation}
t_{\perp}\{T<T^*\}=0.
\end{equation}
{\it This produces a jump in the
electron hopping amplitude}. This can be interpreted as a consequence of
the disappearance of the electron states on chain 1. At $T=T^*$, in 
contrast chain 1 exhibits
preformed bosonic pairs due to the occurrence of the spin gap. In consequence,
one expects the vanishing of the single-particle transverse susceptibility
$\chi_{\perp,s}$\cite{suscep} (See Ref.\cite{Tsvelik} page 224).
This reflects the suppression of the electron
motion in the transverse direction. Such a phenomenon can also
emerge due to the presence of strong Umklapp scattering (leading to
a charge gap)\cite{KLH}.

On the other hand, the Hamiltonian must be supplemented by extra 
terms (i.e., tunneling processes) 
which are generated in the course of renormalization by expanding
the partition function as a function of $t_{\perp}$\cite{Tsvelik}. For a 
two-chain model, those are precisely known and have been 
classified by Khveshchenko and Rice using the bosonic language\cite{K-M}.
Following the scheme introduced in Ref.\cite{KLH}, we now examine
which tunneling processes are relevant below $T^*$. We will
show that because $K_c^->1$, 
these correspond to an incoming electron from chain 2 being reflected
back as a hole, thereby injecting an additional Cooper pair in chain 1.
This is an Andreev scattering mechanism\cite{deGennes}.
\vskip 0.15cm
{\bf Andreev scattering amplitude.}
Terms containing the spin operator 
$\exp[i\beta\theta_{1s}]$ with $\beta=\sqrt{2\pi}$ may be already
dropped. Indeed, below $T^*$
these terms naturally acquire
a scaling dimension greater than 2 and thus become irrelevant in
the sense of the renormalization procedure. Using the classification
scheme of Ref.\cite{K-M} away from half-filling, one must then keep
the following interchain ``pair-hopping'' terms
\begin{eqnarray}
\Delta H &=& \cos\sqrt{4\pi}\theta_{c}^-\hbox{\Large{\{}}
g_2\cos\sqrt{4\pi}\Phi_{s}^-+g_3\cos\sqrt{4\pi}\Phi_{s}^+\hbox{\Large{\}}}
\\ \nonumber
&+&
g_5\cos\sqrt{4\pi}\Phi_{c}^-\cos\sqrt{4\pi}\Phi_{s}^-.
\end{eqnarray}
To fix the values of the coupling constants for
$T\approx T^*$, one must proceed as follows\cite{KLH}. 
\vskip 0.05cm
Above $T^*$, the couplings $g_i$ with i=(2,3) evolve in a similar manner as
$(z(l)=t_{\perp}(l)/E_F)$
\begin{equation}
\frac{dg_i}{dl}=\lambda g_i-\lambda z^2,
\end{equation}
with
\begin{equation}
\lambda=2-[K_s^{\pm}+\frac{1}{K_c^-}]\approx -(g_c+g_s)/2\pi v_F>0
\end{equation}
being slightly renormalized. As long as the spin backscattering is small,
one can use the exponents $(K_{s}^{\pm}<1)$:
\begin{equation}
K_{s}^{\pm}=\sqrt{K_{1s}K_{2s}}\approx 1+\frac{g_s}{2\pi v_F}=
\frac{1}{2}(K_{1s}+K_{2s}).
\end{equation}
The bare values are given precisely by $z(0)=t_{\perp}/E_F$ and
$g_i(0)=0$\cite{Tsvelik}. Under renormalization, these pair-hopping 
terms then acquire a small but {\it non-zero} value
\begin{eqnarray}
\label{g}
g_{2,3}(l)&=&-\frac{\lambda}{2-\lambda}
z(0)^2\hbox{\Large{[}} \exp(2l)-\exp(\lambda l)
\hbox{\Large{]}}
\\ \nonumber
&\approx& -\frac{\lambda}{2-\lambda}z(l)^2,
\end{eqnarray}
with $z(l)=z(0)\exp l$.
Using Eqs.(\ref{tt}) and (\ref{g}) for $T\rightarrow T^*$, one finally finds
$(|g_{2,3}(T^*)|\ll 1)$
\begin{equation}
g_{2,3}(T^*)\propto -\frac{\{t_{\perp}(T^*)\}^2}{{E_F}^2}
=-\frac{{t_{\perp}}^2}{\{T^*\}^2}\cdot
\end{equation}
The exact prefactor is not of interest here. 

The evolution
of the coupling $g_5$ can be discussed in an identical way. 
Starting with
{\it free (or almost free)} electrons on chain 2, we obtain the 
same equation
as for $g_{2,3}$ with $\lambda$ replaced by\cite{second-order}
\begin{equation}
\gamma=2-[K_s^{-}+K_c^-]\approx (g_c-g_s)/2\pi v_F=0.
\end{equation}
Therefore, one reaches the important conclusion that in
our case $g_5(T)=0$ whatever the 
temperature. Note that in principle, this does
not remain true if (quite strong) 
repulsive interactions are added between unpaired
remaining electrons. This provides both $K_c^-<1$ and 
$K_s^{-}<1$\cite{su(2)}. In that
case, the coupling $g_5$ becomes strongly relevant leading to
another spin-gapped (but {\it \`a priori} not superconducting) fixed 
point\cite{KLH,note}.

When T approaches $T^*$, the spin field
$\Phi_{1s}$ gets locked, 
resulting in $<\cos\sqrt{2\pi}\Phi_{1s}>\ \sim \{T^* a\}^{1/2}$ (we put $a=1$)
and then in the unique pair-tunneling mechanism
\begin{equation}
\label{V}
\Delta H=V\cos\sqrt{4\pi}\theta_{c}^-\cos\sqrt{2\pi}\Phi_{2s}.
\end{equation}
This term is the driving force for the quasi long-range Bose condensation
of the preformed pairs. We stress that in 1D, the (bare) interchain
hopping is sufficient to generate such (particle-particle) 
pair-hopping term, i.e.,
$V=-{t_{\perp}}^2/\{T^*\}^{3/2}$. It is appropriate to write
\begin{equation}
\Delta H=V\hbox{\Large{(}}{\cal O}_{SC}^{1\dagger}
\hbox{\large{(}}\psi_{+2\uparrow}\psi_{-2\downarrow}+\psi_{-2\uparrow}
\psi_{+2\downarrow}\hbox{\large{)}}+{\mathrm H.c.}\hbox{\Large{)}}.
\end{equation}
We immediately 
recognize the boson-mediated Cooper pairing between remaining unpaired
electrons, introduced 
in the context of high-$T_c$ cuprates\cite{GIL}. This also
corresponds to an Andreev reflection\cite{deGennes}.

It is worth to note
that the coupling $V$ does not affect the quantum coherence along
the chains. The charge current of each chain 
$J_j\propto v_F\int dx \nabla\theta_{jc}(x)$
commutes with the process $V$, i.e., $[J_j,\Delta H]=0$. For 
clean and infinitely long chains, the conductivity 
$\sigma_{\parallel}$ remains infinite. 
To prompt for
the question of the (in)coherence in the transverse direction below $T^*$ 
one can examine the transverse 
susceptibility $\chi_{\perp,p}$ induced by the Andreev process (transfer
of pairs). As long as $V$ is small, we can expand the partition function up to 
terms of fourth order in $V$. Using Ref.\cite{Tsvelik} (page 223), 
we find $\chi_{\perp,p}\propto T^{2(d_{\perp,p}-1)}$ 
and $d_{\perp,p}=(K_{2s}/2+1/K_c^-)$\cite{suscep}. 
Near $T^*$, we immediately get $d_{\perp,p}>1$. Thus,
$\chi_{\perp,p}$ is first reduced by decreasing the temperature, 
still reflecting a certain incoherence
between the bosonic pairs and the unpaired electrons. 
At lower temperatures the system converges to a 
superconducting and completely coherent fixed point.
Below $T^*$, the coupling $V$ evolves according to $dV/d\ln\hbox{\large{(}}
\frac{T^*}{T}\hbox{\large{)}}=(2-d_{\perp,p})V$\cite{exponents}.
This will be strongly renormalized at the temperature, 
$T_c\approx T^* V^2 =a{t_{\perp}}^4/\{T^*\}^2\ll T^*$. 
Close to $T_c$, the exponents $1/K_c^-$ and $K_{2s}$ 
drastically decrease [similarly as in Eq.(\ref{13})] resulting in
$d_{\perp,p}\{T_c\}\ll 1$. A strong interchain coherence thus arises
below $T_c$; $\chi_{\perp,p}$ diverges at $T=0$. 
The proximity effect results in
 a fluid of coherent bosons with (almost) long-range pairing. 

As in Ref.\cite{EK}, the pre-existing 
spin gap is conveyed to the Fermi gas. 
\vskip 0.08cm
{\bf Superconductivity at} ${\mathbf T=0.}$
The spin field $\Phi_{2s}$ gets massive [See Eq.(22)], so the 
spin fluctuations now contribute a multiplicative constant to all 
the correlation functions.
The SC phase below $T_c$ thus
becomes only a property of the charge degrees of freedom. Here, 
the Andreev reflection imposes that the field $\theta_c^-$ gets locked
(for all $x$), i.e., $\theta_{1c}(x)=\theta_{2c}(x)$.
Only the superfluid phase $\theta_c^+$ remains
massless, thereby producing strong superconductivity. In each chain, one must 
write
$(j=1,2)$
\begin{equation}
\sqrt{2}\theta_{jc}=\frac{1}{\sqrt{2}}\hbox{\large{\{}}\theta_{1c}+\theta_{2c}
\hbox{\large{\}}}
\approx \theta_c^+.
\end{equation}
Using Eq.(\ref{cor}), this gives
\begin{equation}
{\cal O}^j_{SC}\propto\exp(-i\sqrt{\pi}\theta_{c}^+).
\end{equation}
The two chains become obviously phase-coherent and 
the SC correlation functions decay very slowly with $x$,
\begin{equation}
<{\cal O}^{j\dagger}_{SC}(x){\cal O}^j_{SC}(0)>\ \propto 
x^{-1/(2K_{c}^+)}.
\end{equation}
(The exponent is in contrast with the one of Ref.\cite{EK} at Page 6131; 
Our $T_c$ corresponds to their $T^*_2$). 
This exemplifies the almost
Bose condensation of preformed pairs
due to the exchange of fermions
even though a phase order is not strictly possible in 1D, i.e.,
$<{\cal O}^j_{SC}(x)>\ =0$. 

This behavior is reminiscent of a conventional two-leg ladder 
material (our chains behave as bands of the symmetric two-leg
ladder with repulsive interactions)\cite{Fisher,Tsvelik,KLH}. 
We deduce that the pairing susceptibility diverges
approximately as $\chi_{SC}\propto T^{-3/2}$ (See Fig. 2).
Each chain is now characterized by the same superflow. Since
$\theta_c^-\approx 0$, this indeed produces $J_1\approx J_2$. The
density of Cooper pairs in each chain $n_j$ fluctuates strongly, i.e.,
$(n_1-n_2)\propto \partial_x\Phi_c^-$. Finally, the CDW fluctuations 
can develop only at very short distances\cite{CDW}.
To conclude, we have introduced
an asymmetric two-leg ladder system which allows to rigorously investigate
the (almost) Bose condensation of
preformed bosonic pairs in the vicinity
of unpaired fermions discussed in the
context of high-$T_c$ cuprates. The Andreev scattering mechanism has been 
derived properly 
from the fermionic model (above $T^*$). 

We already like to push forward the fact that this asymmetric model
can be generalized to preformed bosonic pairs having an
approximate d-wave pairing and a very small spectral weight (and coexisting
with chain(s) of unpaired fermions), e.g., 
by investigating asymmetric three-leg ladder 
systems\cite{unpublished,3-band,EKZ}. 

{\bf Acknowledgment.}
We thank Lev Ioffe for inspiring us this work, and T.Maurice Rice for 
discussions on superconductivity with preformed pairs.

\begin{figure}[ht]
\centerline{\epsfig{file=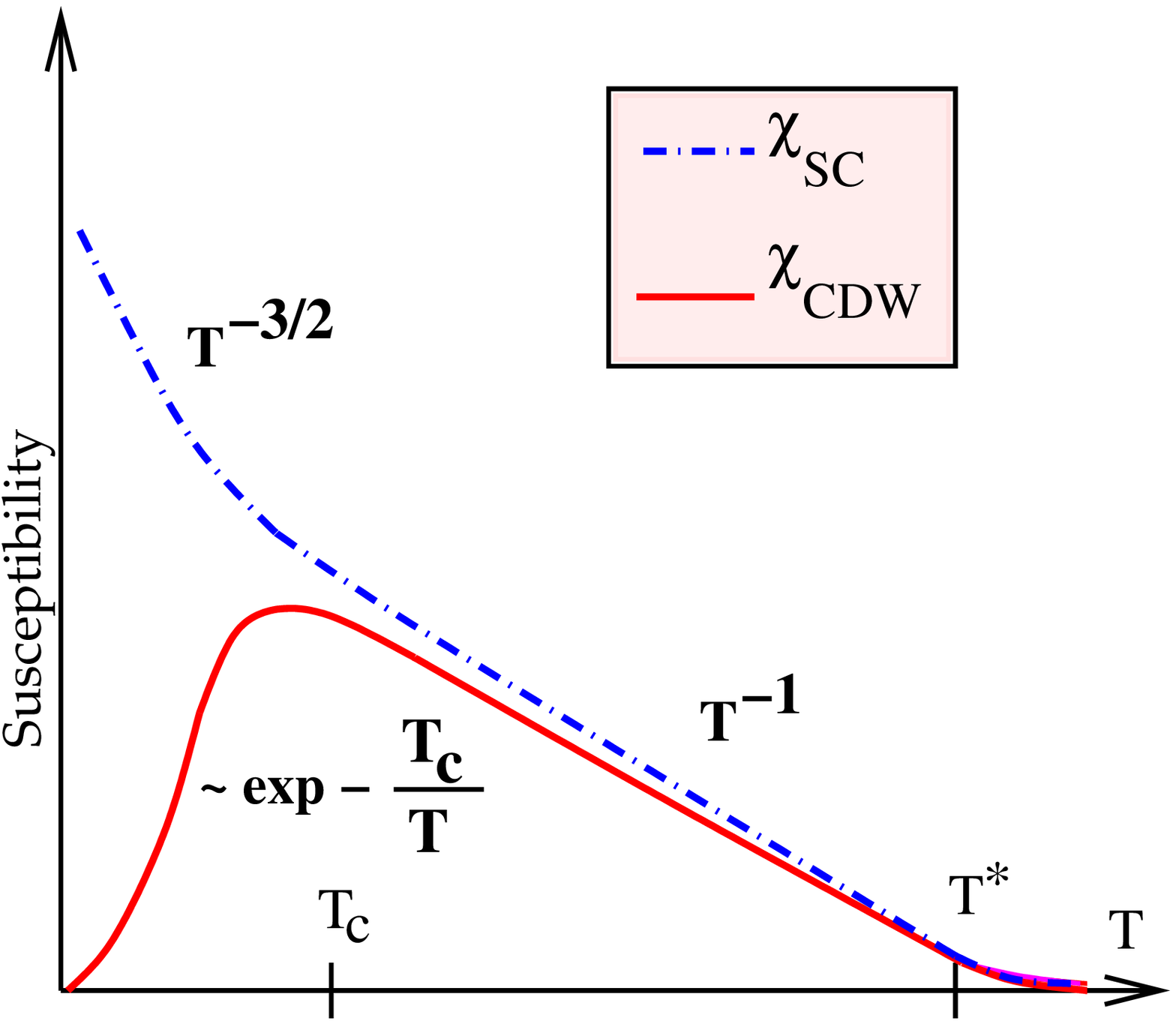,angle=0.0,height=6.5cm,width=7.8cm}}
\vskip 0.3cm
\caption{Schematic vue of the CDW and SC susceptibilities as a function of
the temperature. In general, the corresponding susceptibilities vary as 
$T^{\nu-2}$ (The associated correlation functions decrease as $x^{-\nu}$). 
Below $T_c$, the CDW correlation in contrast exhibits an exponential decay 
resulting in the complete vanishing of the $(2k_F)$ 
CDW susceptibility at $T=0$. We have approximated $(K_{1c};K_{c}^+)\approx 1$.} 
\end{figure}


\begin{references}
\bibitem{velocity} We neglect corrections to the Fermi velocity for 
chain 1.
\bibitem{suscep} We have $\chi_{\perp,s}\propto
T\frac{\partial^2}{\partial{t_{\perp}}^2}\ln{\cal Z}_{\perp}$ 
and $\chi_{\perp,p}\propto
T\frac{\partial^2}{\partial V^2}\ln{\cal Z}_{\perp}$ 
where ${\cal Z}_{\perp}$ is the
$\perp$-part of the grand partition function.
\bibitem{second-order} Taking into account second order corrections to
$K_{c}^-$ and $K_s^-$ would not change the further conclusions.
\bibitem{su(2)} For repulsive interactions in chain 2, the 
rotation invariance implies
that for $T\ll E_F$: $K_{2s}=1, K_{s}^{\pm}<1$.
\bibitem{note} The term $g_5$ produces
a coupling ${\cal O}^{2\dagger}_{CDW}\psi^{\dagger}_{+1\alpha}\psi_{-1\alpha}$.
This would lead to a spin-gapped CDW ground state $(\lambda\approx 0)$.
\bibitem{exponents} For $T\leq T^*$, in Eqs.(16) and (18) one must 
formally replace
$g_i\rightarrow V$ and
equate $z(l)=0$, $K_{1s}(l)=0$.
\bibitem{CDW} Similarly, we obtain:
\begin{eqnarray}
{\cal O}^j_{CDW}&\propto&\exp(-i\sqrt{2\pi}\Phi_{jc})\\ \nonumber
&=&\exp(-i\sqrt{\pi}\Phi_c^+)\exp(\mp i\sqrt{\pi}\Phi_c^-).
\end{eqnarray}
The two-point correlation functions of the disordered
field $\Phi_c^-$ decay exponentially
(its dual field is pinned), therefore $<{\cal O}^{j\dagger}_{CDW}(x)
{\cal O}^j_{CDW}(0)>\ \propto\exp(-x T_c)$ (in contrast with Ref.\cite{EK}).
\bibitem{3-band} But the stability of the 
``C2S1'' phase in the symmetric 3-leg Hubbard ladder (or 3-band Hubbard model)
of Ref.\cite{Maurice-ladder} is clear. Starting from the half-filled case,
the unpaired fermions of band 2 have been found to be 
completely decoupled from bands 1 and 3 (which contain the bosonic pairs). 
Adding (by hand) a scattering term of the form 
\begin{equation}
V\hbox{\Large{(}}
({\cal O}_{SC}^{1\dagger}+{\cal O}_{SC}^{3\dagger})
\hbox{\large{(}}\psi_{+2\uparrow}\psi_{-2\downarrow}+\psi_{-2\uparrow}
\psi_{+2\downarrow}\hbox{\large{)}}+{\mathrm H.c.}\hbox{\Large{)}}
\end{equation}
would not affect the fixed point. The
bosonic pairs in bands 1 and 3 indeed obey 
${\cal O}^1_{SC}\approx -{\cal O}^3_{SC}$ and the unpaired fermions
are located exactly on the diagonals of the 2D Fermi surface, i.e.,
$\Delta H\propto (V-V)=0$.
\end{references}
\end{document}